\documentclass[twocolumn]{aastex63}
\usepackage{threeparttable}
\usepackage[caption=false]{subfig}
\usepackage{graphicx}
\usepackage{multirow}
\usepackage{graphicx}
\usepackage{booktabs}
\usepackage{amsmath,amssymb}
\usepackage[T1]{fontenc}
\usepackage{color}
\usepackage{epstopdf}
\graphicspath{{./}{figures/}}

\begin{document}

\title{Study of $\gamma$-ray Emission from a Compact Radio Galaxy with the Fermi Large Area Telescope}

\correspondingauthor{Hai-Ming Zhang \& Jin Zhang}
\email{hmzhang@nju.edu.cn; j.zhang@bit.edu.cn}

\author[0009-0006-4551-8235]{Tan-Zheng Wu}
\affiliation{School of Physics, Beijing Institute of Technology, Beijing 100081, People's Republic of China; j.zhang@bit.edu.cn}

\author[0009-0000-6577-1488]{Yu-Wei Yu}
\affiliation{School of Physics, Beijing Institute of Technology, Beijing 100081, People's Republic of China; j.zhang@bit.edu.cn}

\author[0000-0002-4789-7703]{Ying-Ying Gan}
\affiliation{School of Physics, Beijing Institute of Technology, Beijing 100081, People's Republic of China; j.zhang@bit.edu.cn}

\author[0000-0001-6863-5369]{Hai-Ming Zhang\dag}
\affiliation{School of Astronomy and Space Science, Nanjing University, Nanjing 210023, People's Republic of China; hmzhang@nju.edu.cn}

\author[0000-0003-3554-2996]{Jin Zhang\dag\dag}
\affiliation{School of Physics, Beijing Institute of Technology, Beijing 100081, People's Republic of China; j.zhang@bit.edu.cn}

\begin{abstract}

The radio galaxy PKS 1007+142 is classified as a compact steep-spectrum source (CSS) and belongs to the class of young Active Galactic Nuclei (AGNs). In this paper, we investigate the $\gamma$-ray emission from this CSS by conducting a comprehensive analysis of the 15 yr Fermi Large Area Telescope (Fermi-LAT) observation data. The Fermi-LAT latest Source Catalog, 4FGL-DR4, includes an unassociated $\gamma$-ray source, 4FGL J1010.0+1416, located at 0.24$\degr$ away from the radio position of PKS 1007+142. Using the 15 yr Fermi-LAT observation data, we re-estimate the best-fit position of the $\gamma$-ray source and find that PKS 1007+142 is in close proximity to the $\gamma$-ray source and falls within its 68\% error circle. Therefore, we conclude that PKS 1007+142 is the most plausible counterpart to the unassociated LAT source with a detection test statistics (TS) $\sim$43.4 ($\sim 6.6\sigma$). PKS 1007+142 exhibits a steep power-law spectrum in the 0.1--300 GeV band, with a photon spectral index ($\Gamma_{\gamma}$) of $2.86\pm0.17$. The average flux in the considered time interval is $\rm (2.14\pm0.34)\times10^{-12}\ erg\ cm^{-2}\ s^{-1}$. Comparing PKS 1007+142 with other $\gamma$-ray emitting AGNs in both the $L_{\gamma}-\Gamma_{\gamma}$ and $L_\gamma-L_{\rm 1.4GHz}$ planes, it shows a softer $\gamma$-ray spectrum and lower luminosity compared to other $\gamma$-ray emitting CSSs. Furthermore, the possible origins of $\gamma$-ray in PKS 1007+142 are also discussed. 

\end{abstract}

\keywords{galaxies: active---galaxies: jets---radio continuum: galaxies---gamma-rays: galaxies}

\section{Introduction}
\label{sec:intro}

Exploring the $\gamma$-ray emitting sources in the universe is crucial for understanding the high-energy physical processes and the history of galaxy evolution \citep{fanti1995compact,o1997constraints}. Radio-loud Active Galactic Nuclei (RL-AGNs) are one of the main class of the $\gamma$-ray emitting extragalactic sources \citep[see 4LAC-DR2,][]{lott2020fourth}. Compact radio sources (CRSs), which are a sub-class of RL-AGNs, have a projected linear size smaller than about 20 kiloparsec \citep [kpc,][]{o1998compact,fanti1990nature}. They tend to be located at elliptical galaxies \citep{de2000hubble}, similar to radio galaxies and blazars \citep{wold2000clustering,ramos2013environments}, indicating that CRSs are in a similar environment to other large radio sources. There are three main explanations for the small size of CRSs: the relatively dense environment around their jets results in a frustrated jet; they are thought to be the progenitors of the typical radio galaxies with large-scale jets \citep{readhead1996statistics,an2011vlbi}, being the young sources in the early stage of the AGN evolution; recently, \citet{readhead2023evolution} suggested that CRSs should be characterized as short-lived, as opposed to young, and their jets cannot be transported to a large scale \citep[see also][]{readhead1993evidence,reynolds1997intermittent}. The last point is similar to the explanation of the transient or episodic sources \citep{o2021compact,lister2020txs,2021RAA....21..201G,2022ApJ...939...78G}. Anyway, the distinct traits between CRSs and other jetted-AGNs provide crucial insights into the formation and evolution of relativistic jets in AGNs.

Up to now, about a dozen $\gamma$-ray emitting CRSs have been identified using the Fermi Large Area Telescope (Fermi-LAT) observations \citep[e.g.,][]{migliori2016first,principe2020ngc,principe2021gamma,abdollahi2020fermi,2020ApJ...899....2Z,lister2020txs,2022ApJ...927..221G,2021RAA....21..201G,2022ApJ...939...78G,2024RAA....24b5018G}. However, the origin of their $\gamma$-ray emission is not so clear. It is speculated that the $\gamma$-ray emission of the quasar-type CRSs originates from the jet in the core region while the $\gamma$-ray emission of galaxy-type CRSs is relatively weak and insignificant variation, and it is dominated by the emission from the extended radio lobes \citep[e.g.,][]{stawarz2008momentum,migliori2013jet,principe2021gamma,2024RAA....24b5018G}. The radiation mechanism of $\gamma$-rays for these CRSs is normally the inverse Compton scattering of the relativistic electrons in the core-jet or extended lobes \citep[e.g.,][]{stawarz2008momentum,migliori2013jet,2020ApJ...899....2Z,2022ApJ...927..221G,2022ApJ...939...78G}. The thermal bremsstrahlung \citep{kino2007extragalactic,kino2009new} and the hadronic radiation processes in mini radio lobes on parsec scale may also result in two bumps at $\sim$1 MeV and $\sim$1 GeV in the broadband spectral energy distributions (SEDs) of CRSs \citep{kino2011mini}. There are few CRSs detected at the $\gamma$-ray band so that adding even only one to the sample would be beneficial for revealing their $\gamma$-ray radiation property.

The low-luminosity compact object PKS 1007+142, located at a redshift of $z$ = 0.213 \citep{2012ApJS..203...21A},  is characterized as a `double-lobed' object based on preliminary radio images at 1.6 GHz  \citep{2009AN....330..210K,2010MNRAS.408.2261K}. It is classified as a medium symmetric object (MSO) with a projected linear size of 3.3 kpc and is also identified as a GHz-peak spectrum source due to its peak frequency in the radio band falling within the range of 0.5--2 GHz \citep{principe2021gamma}. The $\gamma$-ray source 4FGL J1010.0+1416 (TS = 31, $\Gamma=2.73\pm0.19$), initially reported in the 4FGL-DR2 \citep{2020arXiv200511208B}, is located in the vicinity of PKS 1007+142, however, no association between PKS 1007+142 and 4FGL J1010.0+1416 was reported in the 4FGL-DR2. By analysing the 11.3 yr Fermi-LAT observational data, \citet{principe2021gamma} suggested that PKS 1007+142 is a $\gamma$-ray emitting source (TS = 31, $\Gamma=2.56\pm0.18$). The LAT best-fitting position of PKS 1007+142 obtained by \citet{principe2021gamma} is within a distance of less than 0.2$\degr$ from the $\gamma$-ray source 4FGL J1010.0+1416. In this paper, we deeply investigate the potential association between PKS 1007+142 and 4FGL J1010.0+1416 by analyzing the 15 yr Fermi-LAT observation data. Then, we examine the $\gamma$-ray emission characteristics of PKS 1007+142 by comparing with other $\gamma$-ray emitting AGNs.

\section{Fermi-LAT Data Analysis and Results}

\subsection{Data Selection}

The Pass 8 data of the PKS 1007+142 region, covering 15 yr (MJD 54683--60160, from 2008 August 4 to 2023 August 4), are extracted from Fermi Science Support Center\footnote{https://fermi.gsfc.nasa.gov/ssc/data/access/} for our analysis. We select the region of interest (ROI) centered at the radio position (R.A.= 152.48$\degr$, Decl.= 14.03$\degr$) of PKS 1007+142 \citep{1996AJ....111.1945D} with a radius of 15$\degr$. The publicly available software \textit{Fermipy} (version v1.1)\citep{Wood2017} and \textit{Fermitools} \citep[version 2.2.0,][]{2019ascl.soft05011F} are used to perform the data analysis with the binned maximum-likelihood method. The $\gamma$-ray events in the energy range of 0.1--300 GeV are selected with a standard data quality selection criteria of “(DATA\_QUAL\textgreater0)\&\&(LAT\_CONFIG==1)”. We binned the data with a pixel size of $0.1\degr$ and twelve energy bins per decade. To reduce the $\gamma$-ray contamination from the Earth limb, we implemented a stringent selection criterion by focusing on event types that exhibit the most favorable point spread function (PSF), following the analysis conducted by \citet{principe2021gamma}. We exclude events with zenith angles exceeding 85$\degr$ for an energy range of 0.1--0.3 GeV, including photons from PSF0 and PSF1 event types. We also exclude events surpassing 95$\degr$ for an energy range of 0.3--1 GeV, as well as photons belonging to the PSF0 event type. Finally, we use all events greater than 1 GeV with zenith angles less than 105$\degr$. The event class $P8R3\_SOURCE$ (“evclass=128”), and corresponding instrument response functions (IRFs) P8R3\_SOURCE\_V3 \citep{bruel2018fermi} are used in our analysis. 

\subsection{Residual TSmap Test}

\begin{figure}
 \centering
   \includegraphics[angle=0,scale=0.40]{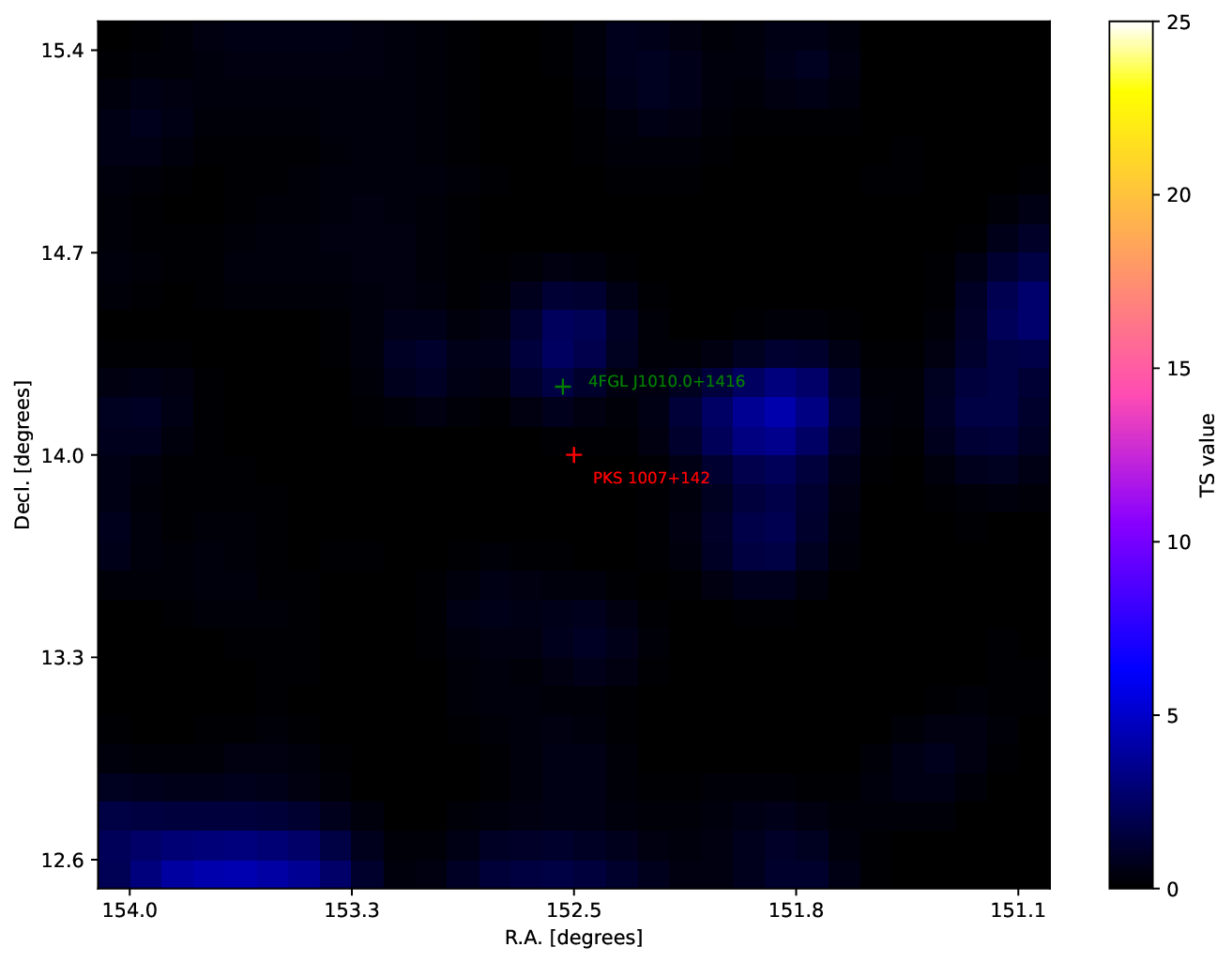}
   \includegraphics[angle=0,scale=0.40]{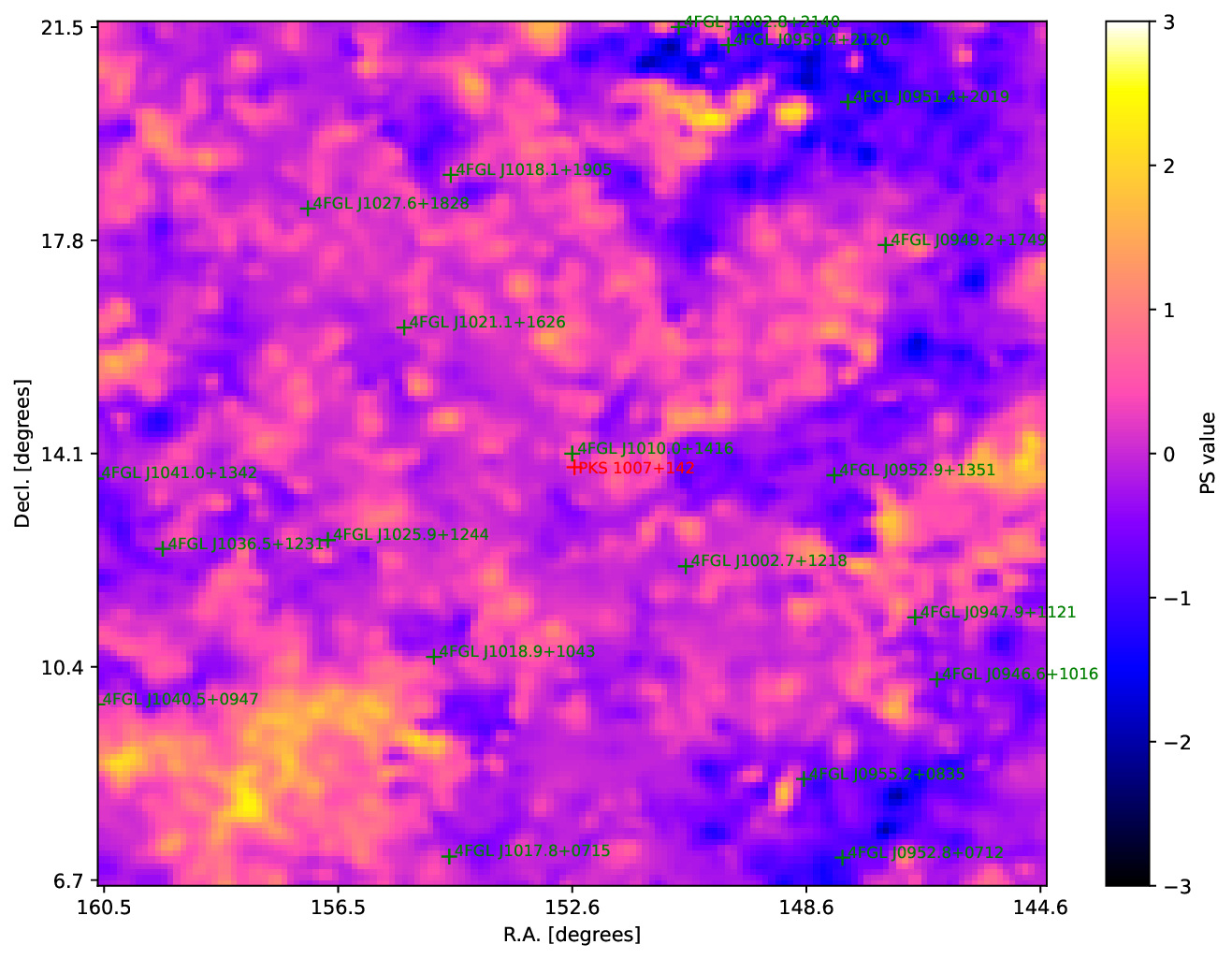}
\caption{\emph{Left panel}: The $3\degr\times3\degr$ residual TS map of the ROI centered at the radio position of PKS 1007+142 in energy band of 0.1--300 GeV generated by \textit{gttsmap} from \textit{Fermitools v2.2.0}. \emph{Right panel}: The $15\degr\times15\degr$ PS map of the ROI centered at the radio position of PKS 1007+142 generated by \textit{gtpsmap} \citep{2021A&A...656A..81B}. The green crosses represent the positions of the sources in the 4FGL-DR4 within ROI, while the red cross represents the radio position of PKS 1007+142.}\label{resTSmap}
\end{figure}

Since the selected data covers a span of 15 years, which differs from the Fermi-LAT latest Source Catalog (4FGL-DR4, \citealp{ballet2023fermi,2022ApJS..260...53A}), we first conduct a new background source test for the ROI. The maximum likelihood test statistic (TS) is used to estimate the significance of $\gamma$-ray signals, where $\rm{ TS = 2log(\frac{\mathcal{L}_{src}}{\mathcal{L}_{null}})}$, $\rm{\mathcal{L}_{src}}$ and $\rm{\mathcal{L}_{null}}$ are the likelihood values for the background with or without the source. We set a background model (BGM) of known $\gamma$-ray sources in the ROI, including the isotropic background model iso\_P8R3\_V3\_v1.txt and the diffuse Galactic interstellar emission with the parameterized model gll\_iem\_v07.fits, as well as all the $\gamma$-ray sources listed in the 4FGL-DR4. The normalization of the isotropic background emission and the diffuse Galactic interstellar emission, together with the normalization and spectral parameters of the $\gamma$-ray point sources within a 5$\degr$ radius centered on PKS 1007+142, are set to be free in the fitting procedure. After subtracting the BGM, we generate the $3\degr\times3\degr$ residual TS map of the ROI centered at the radio position of PKS 1007+142, as depicted in the left panel of Figure \ref{resTSmap}. The maximum TS value in the residual TS map is 8.1, indicating that no new $\gamma$-ray source is found. We do not find any excess $\gamma$-ray signal in the residual TS map after subtracting all the sources already included in the 4FGL-DR4. 

For crosscheck, we employ a fast and reliable tool called \textit{gtpsmap} \citep{2021A&A...656A..81B} to assess the goodness-of-fit of Fermi-LAT data. It is a python script that compares the 3D map of the data with the 3D map of the model by computing a point source (PS) map. This tool adopts a log-likelihood approach and defines the following random variable: $L = -\sum_{k} \log \mathcal{P}(x_k, m_k)$, where $\mathcal{P}$ represents the Poisson probability, $x_k$ denotes independent random Poisson variables with mean $m_k$, which is the spatially integrated number of model counts in the spectral bin $k$. The data/model deviation estimator, PS, is defined as: $\lvert PS \rvert = -\log_{10}(\text{p-value})$. The p-value is the integral of the probability distribution function (pdf) of $L$ above $L_{data} = -\sum_{k} \log \mathcal{P}(n_k, m_k)$, which is obtained with the data integrated count spectra $n_k$ \citep{2021A&A...656A..81B}. Using the \textit{gtpsmap} tool, we generate the $15\degr\times15\degr$ residual map of ROI centered at the radio position of PKS 1007+142, as shown in the right panel of Figure \ref{resTSmap}. Within a 5-degree radius centered on PKS 1007+142, there is no significant $\gamma$-ray emission (PS \textless \ 3 corresponding to $\sigma$ \textless \ 3.29, \citealp{2021A&A...656A..81B}).

\subsection{Detection of $\gamma$-rays from PKS 1007+142}

\begin{figure}
 \centering
   \includegraphics[angle=0,scale=0.45]{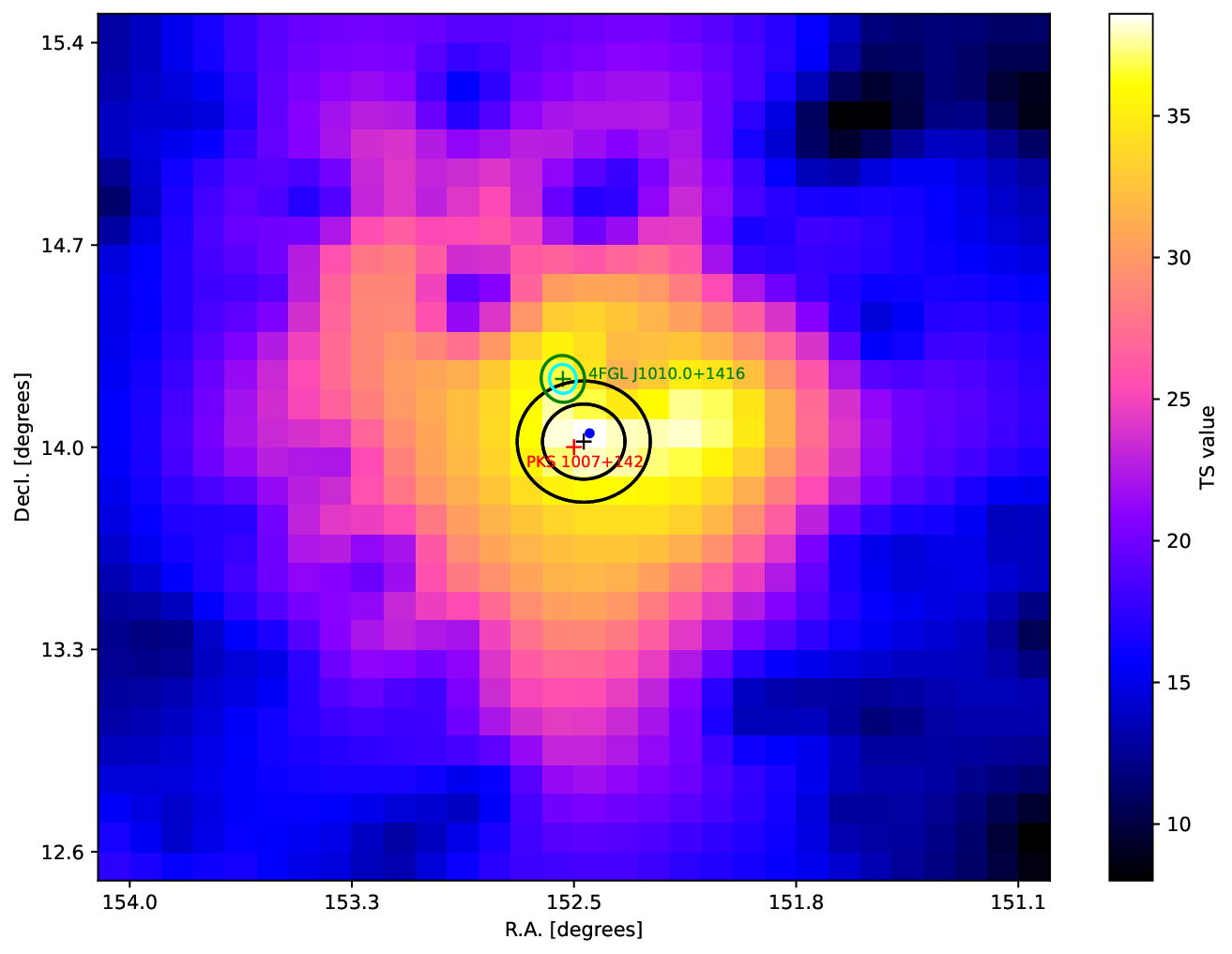}
\caption{The $3\degr\times3\degr$ TS map of PKS 1007+142 in the energy band of 0.1--300 GeV. The black cross represents the re-estimated best-fit position of $\gamma$-ray source with the 15 yr Fermi-LAT observation data, while the two black circles are its corresponding 68\% and 95\% containment regions. The red and green crosses represent the radio position of PKS 1007+142 and the position of 4FGL J1010.0+1416 in the 4FGL-DR4, respectively. The cyan and green contours, centered at the location of 4FGL J1010.0+1416, indicate its localisation uncertainty and its 95\% containment in the 4FGL-DR4, respectively. The blue dot represents the LAT best-fitting position of PKS 1007+142 reported in \citet{principe2021gamma}.}\label{tsmap}
\end{figure}

Although there is no excess residual $\gamma$-ray signal in the vicinity of PKS 1007+142, there is a LAT source, 4FGL J1010.0+1416, located 0.24$\degr$ away from the radio position of PKS 1007+142. 4FGL J1010.0+1416 is the closest $\gamma$-ray source to PKS 1007+142 in the 4FGL-DR4, and it has no associated counterpart reported in the 4FGL-DR4 \citep{ballet2023fermi,2022ApJS..260...53A}. We thus investigate the association between 4FGL J1010.0+1416 and PKS 1007+142. Since our selected data includes an additional year compared to the data used in the 4FGL-DR4, we re-estimate the location of 4FGL J1010.0+1416. The best-fit position of the $\gamma$-ray source is re-estimated using the current BGM (in this case, 4FGL J1010.0+1416 is not subtracted as a background source) and the \textit{Fermipy} tool with the 15 yr Fermi-LAT observation data. The re-estimated best-fit position is located at [R.A.= 152.45$\degr\pm0.08\degr$, Decl.= 14.05$\degr\pm0.09\degr$] with an uncertainty radius of 0.13$\degr$ at the 68\% confidence level, as depicted in Figure \ref{tsmap}. The radio position (R.A.= 152.48$\degr$, Decl.= 14.03$\degr$) of PKS 1007+142 \citep{1996AJ....111.1945D} and the position of 4FGL J1010.0+1416 (R.A.= 152.52$\degr\pm0.04\degr$, Decl.= 14.27$\degr\pm0.05\degr$) in the 4FGL-DR4 \citep{ballet2023fermi,2022ApJS..260...53A}, along with the best-fit position (R.A.= 152.43$\degr\pm0.05\degr$, Decl.= 14.08$\degr\pm0.06\degr$) reported in \citet{principe2021gamma}, are also presented in Figure \ref{tsmap}. The source 4FGL J1010.0+1416 is located at a distance of 0.23$\degr$ from the re-estimated best-fit position, lying beyond the 95\% error circle of the re-estimated best-fit position.  It can be observed that the radio position of PKS 1007+142 is in close proximity to the re-estimated best-fit position of the $\gamma$-ray source, with a distance between them of only $\sim0.03\degr$. These findings suggest a spatial association between PKS 1007+142 and the $\gamma$-ray source with the re-estimated best-fit position. The re-estimated best-fit position yields TS$\sim$43.4 ($\sim6.6\sigma$) for the $\gamma$-ray source. The location results are summarized in Table \ref{location}.

The $\gamma$-ray source 4FGL J1010.0+1416 was detected and reported as a new $\gamma$-ray source in the 4FGL-DR2 \citep{2020arXiv200511208B}. However, it is worth noting that in the subsequent papers of 4FGL-DR3 and 4FGL-DR4, no re-estimation of the position of this source has been conducted\footnote{For details see in \url{https://fermi.gsfc.nasa.gov/ssc/data/access/}}. To crosscheck, we repeated the analysis following the 4FGL-DR2 paper \citep{2020arXiv200511208B}; considering the statistical error, our results are consistent with those of the 4FGL-DR2 paper. The results obtained from a 15 yr dataset demonstrate the advantages of dedicated analysis and prolonged exposure.

\begin{figure}
 \centering
   \includegraphics[angle=0,scale=0.33]{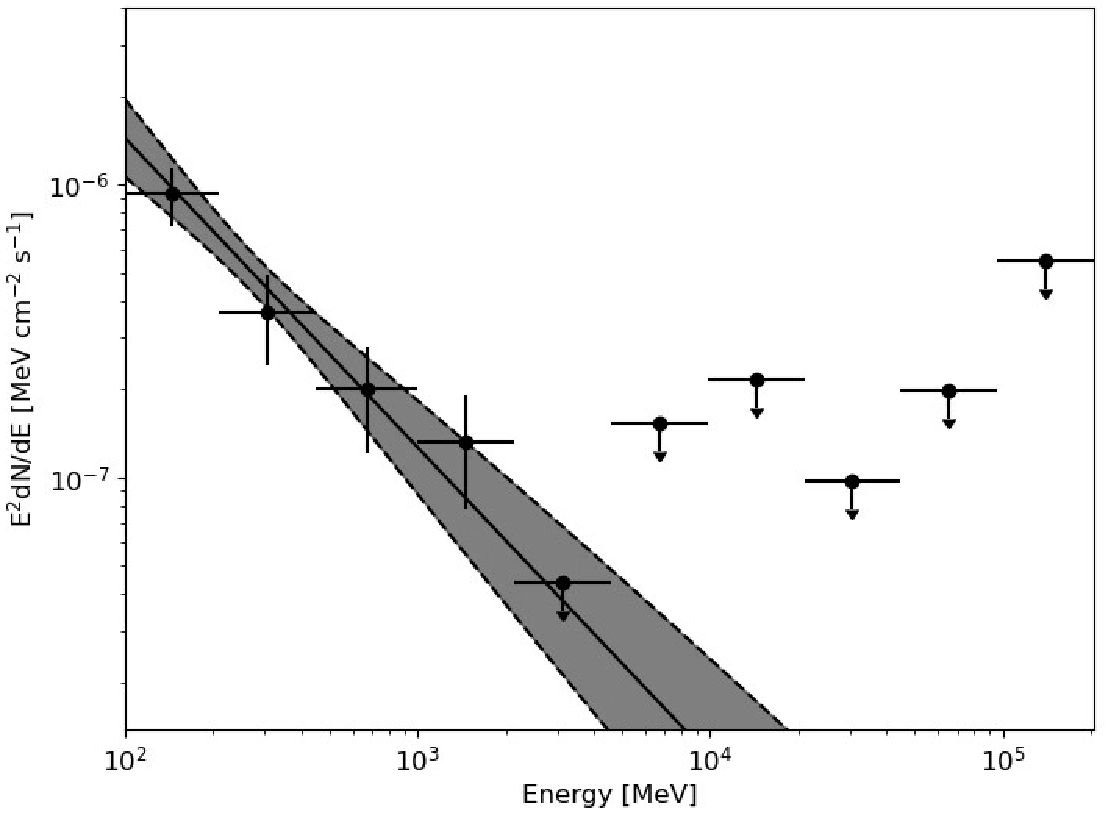}
   \includegraphics[angle=0,scale=0.34]{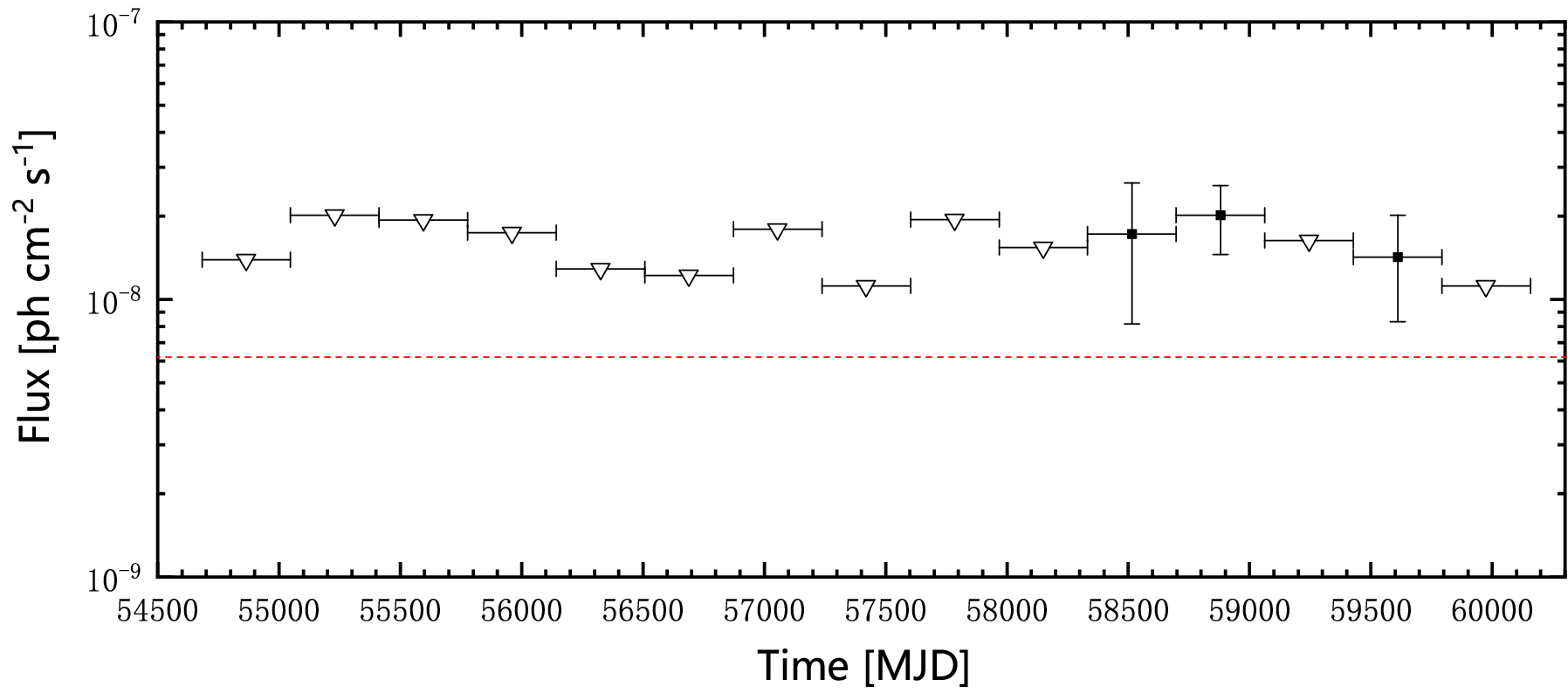}
\caption{\emph{Left panel}: The average spectrum of PKS 1007+142 in the 0.1--300 GeV band derived with the 15 yr Fermi/LAT observation data. The black solid line represents the fitting result with a power-law function, while the grey shaded region indicates the 1$\sigma$ uncertainty. If TS$\le$4, an upper-limit is presented for that energy bin. \emph{Right panel}: The long-term Fermi-LAT light curve of PKS 1007+142 in time bins of 1 yr. The opened triangles indicate TS$\le$9 for that time bin and the horizontal red dotted line represents the 15 yr average flux, i.e., $(6.20\pm1.27)\times10^{-9}$ ph cm$^{-2}$ s$^{-1}$.}\label{spect-LC}
\end{figure}

The photon spectrum of PKS 1007+142 in the 0.1--300 GeV band can be described by a power-law spectral function, specifically $dN(E)dE=N_0(E/E_0)^{-\Gamma_{\gamma}}$, where $N(E)$ represents the photon distribution as a function of energy and $\Gamma_{\gamma}$ is the photon spectral index. We obtain $\Gamma_\gamma =2.86\pm0.17$, $\rm N_0=(2.72\pm0.57)\times10^{-13}\ cm^{-3}$, and $E_0 = 827.43$ MeV. The integrated spectrum and light curve over a period of 15 yr are shown in Figure \ref{spect-LC}. The normalization of the two diffuse sources in the BGM, along with the normalization and spectral parameters of the $\gamma$-ray point sources within a 5$\degr$ radius centered on PKS 1007+142, are set as free variables to generate detection points for the light curve. The upper limits in the light curve are obtained by fixing all parameters except for the normalization parameter of the target source, where the photon spectral index of the target source is fixed at $\Gamma_{\gamma}=2$. The average flux over this period in the 0.1--300 GeV range is $\rm (2.14\pm0.34)\times10^{-12}\ erg\ cm^{-2}\ s^{-1}$. Using the same method applied in the 4FGL catalog \citep{abdollahi2020fermi}, we calculate the variability index (TS$_{\rm var}$) to quantify the variability of this $\gamma$-ray source, yielding a value of TS$_{\rm var}\sim7.89$, which corresponds to a confidence level of 2.1$\sigma$. The flux values/upper limits, uncertainties, and TS values for each time bin are reported in in Table \ref{flux_table}.

To further investigate the association of the $\gamma$-ray emission to the radio source PKS 1007+142, we re-analyze the 15 yr Fermi-LAT data starting from the coordinates of 4FGL J1010.0+1416 and PKS 1007+142's radio position, respectively. As a crosscheck, TS$\sim$39.7 is obtained using the LAT source position, while it yields TS$\sim$41.5 using the radio source position. Comparing the TS value of the re-estimated best-fit position with that of radio position, there is no clear indication to prefer one over the other. Therefore, we investigate multi-wavelength observations to possibly unveil the association between the $\gamma$-ray source and PKS 1007+142.

We examine the 95\% error circle of the re-estimated best-fit position to identify any other potential radio or X-ray counterparts that may be associated with the $\gamma$-ray source. To conduct this search, we utilize the SIMBAD Astronomical Database \citep{2000A&AS..143....9W} to examine the 95\% error circle of the re-estimated best-fit position and identify a total of 47 objects within this region. Apart from the target source and three quasars, the majority of these objects are galaxies and stars. However, no radio or X-ray observational data are available for the three quasars in the \citeauthor{NED} \citeyear{NED}. Hence, PKS 1007+142 is the most possible $\gamma$-ray emitting counterpart within this error circle. Additionally, we employ a Bayesian analysis of 15 yr Fermi-LAT observation data using the \textit{gtsrcid} tool \citep{abdo2010fermi} to estimate the probability of association between the $\gamma$-ray source and PKS 1007+142. The analysis uncovers a significantly high probability of association, reaching 99.2\%, suggesting that PKS 1007+142 is indeed responsible for the detected $\gamma$-ray emission.

\section{Discussion and conclusion}
\label{sec:dis}

Our analysis of the 15 yr Fermi-LAT observation data resulted in a $>5\sigma$ detection of the $\gamma$-ray source 4FGL J1010.0+1416, for which no association was reported in the 4FGL-DR4 \citep{ballet2023fermi,2022ApJS..260...53A} while an initial detection of was reported in \cite{principe2021gamma}. By re-estimating the best-fit position of this $\gamma$-ray source, we obtained a TS value of 43.4. We found that the radio position of PKS 1007+142 is very close to the re-estimated best-fit position and falls within its 68\% error circle, whereas 4FGL J1010.0+1416 is situated at a spatial distance of $0.23\degr$ from the re-estimated best-fit position, lying outside its 95\% error circle. This suggests a significant detection ($>5\sigma$) of $\gamma$-rays originating from the young radio source PKS 1007+142. PKS 1007+142 presents a soft power-law spectrum in the 0.1--300 GeV band with $\Gamma_{\gamma}=2.86\pm0.17$. This spectral shape is comparatively softer than the one obtained with 11.3 yr of Fermi-LAT data in \cite{principe2021gamma}.

\begin{figure}
 \centering
   \includegraphics[angle=0,scale=0.45]{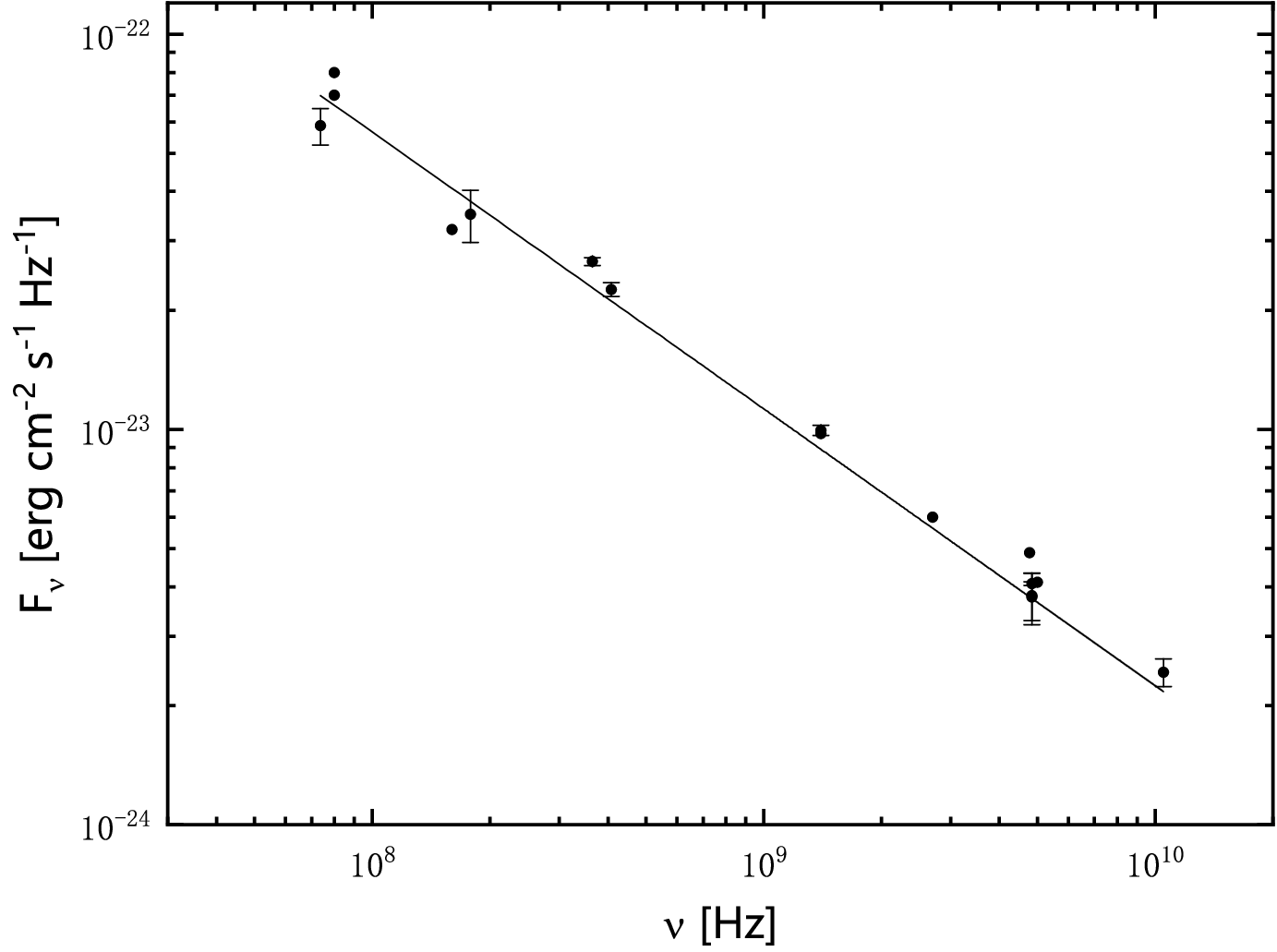}
\caption{The spectrum of PKS 1007+142 in radio band with the power-law fitting line. The data are taken from the NED.}\label{radio}
\end{figure}

The observational data for PKS 1007+142 were collected from various archival multi-wavelength data for this source to investigate its radiation properties, especially the $\gamma$-rays. However, only the flux densities at radio-IR-optical-UV bands\footnote{The observations of Chandra, Swift-XRT, NuSTAR, and XMM-Newton have been checked; however, there are no X-ray observations covering the vicinity of this source.} were obtained from the NASA/IPAC Extragalactic Database (NED)\footnote{\url{https://ned.ipac.caltech.edu/}} and the Space Science Data Center\footnote{\url{https://tools.ssdc.asi.it/SED/}}. Using a power-law function to fit the data at the radio band (Figure \ref{radio}) i.e., $F_{\nu}\propto\nu^{-\alpha}$, we derive a spectral index of $\alpha\sim0.70\pm0.05$. Therefore, it belongs to a steep spectrum radio galaxy, where the radio emission is dominated by optically thin synchrotron radiation in the extended regions. PKS 1007+142 is categorized as a MSO by \citet{principe2021gamma}. Based on its spectral slope of $\alpha\sim0.70\pm0.05$, it is also classified as a compact steep-spectrum source (CSS). The radiation in the IR-optical bands of PKS 1007+142 is primarily dominated by the starlight of its host galaxy and exhibits the significant thermal emission component, which distinguishes it from other seven $\gamma$-ray emitting CSSs \citep{2020ApJ...899....2Z,2022ApJ...927..221G}. Based on the broadband SED fitting, the $\gamma$-rays from three `bona fide' compact symmetric objects (CSOs; PKS 1718–649, NGC 3894, and TXS 0128+554) are suggested to be lobe dominated (\citealt{2022ApJ...941...52S,2024RAA....24b5018G}; see also \citealt{migliori2016first,principe2020ngc}), while the $\gamma$-rays from the two `fake' CSOs (CTD 135 and PKS 1413+135; \citealt{2021RAA....21..201G,2022ApJ...939...78G}; see also \citealt{principe2021gamma}) and seven CSSs \citep{2020ApJ...899....2Z,2022ApJ...927..221G} are core-jet dominated similar to blazars. Due to limited observational data availability, it is impractical to investigate the origin of $\gamma$-rays in PKS 1007+142 using broadband SED fitting based on radiation models, as even a simple one-zone synchrotron-self-Compton homogeneous model requires seven independent parameters \citep{1998ApJ...509..608T}.

\begin{figure}
 \centering
   \includegraphics[angle=0,scale=0.44]{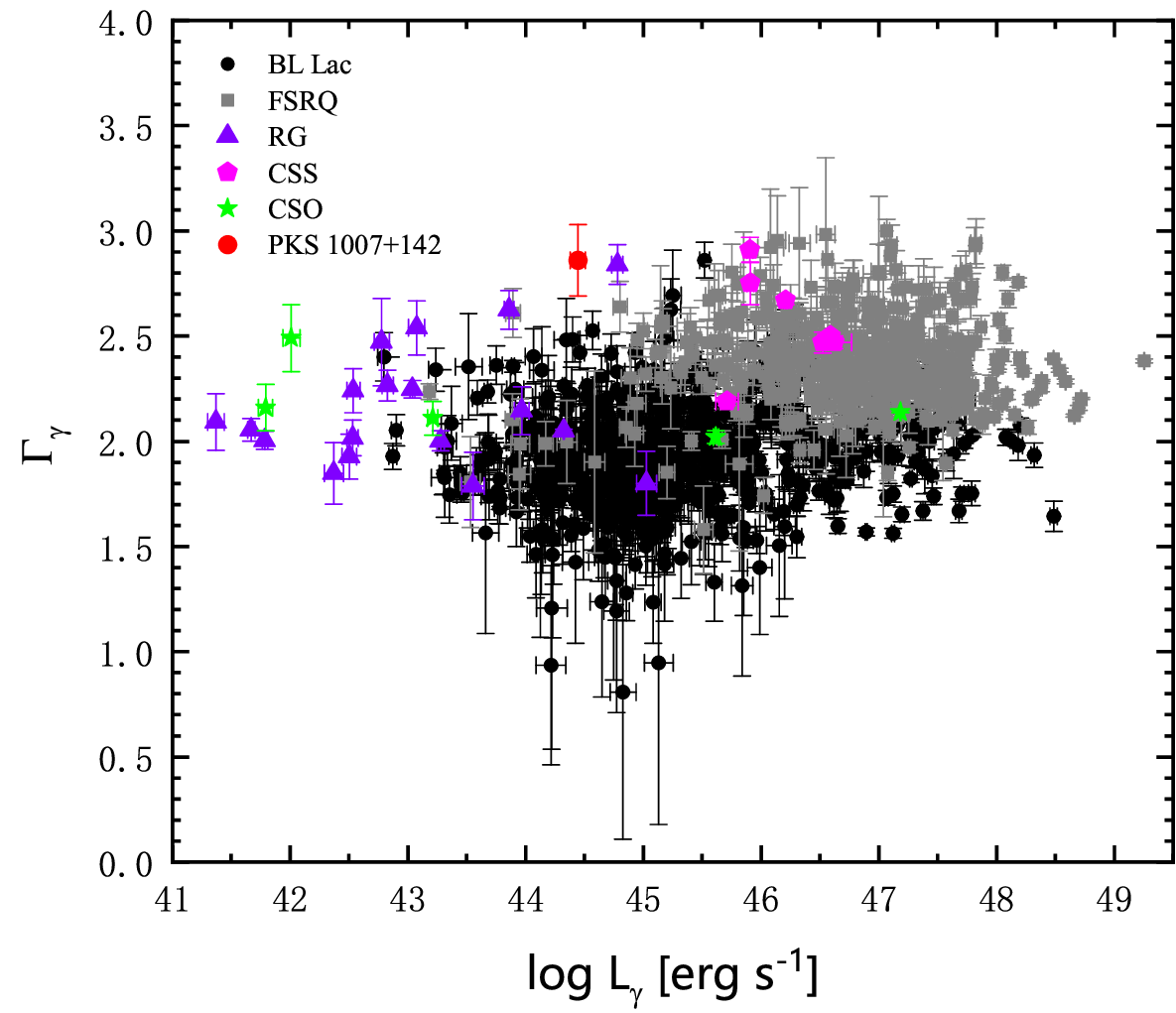}
   \includegraphics[angle=0,scale=0.44]{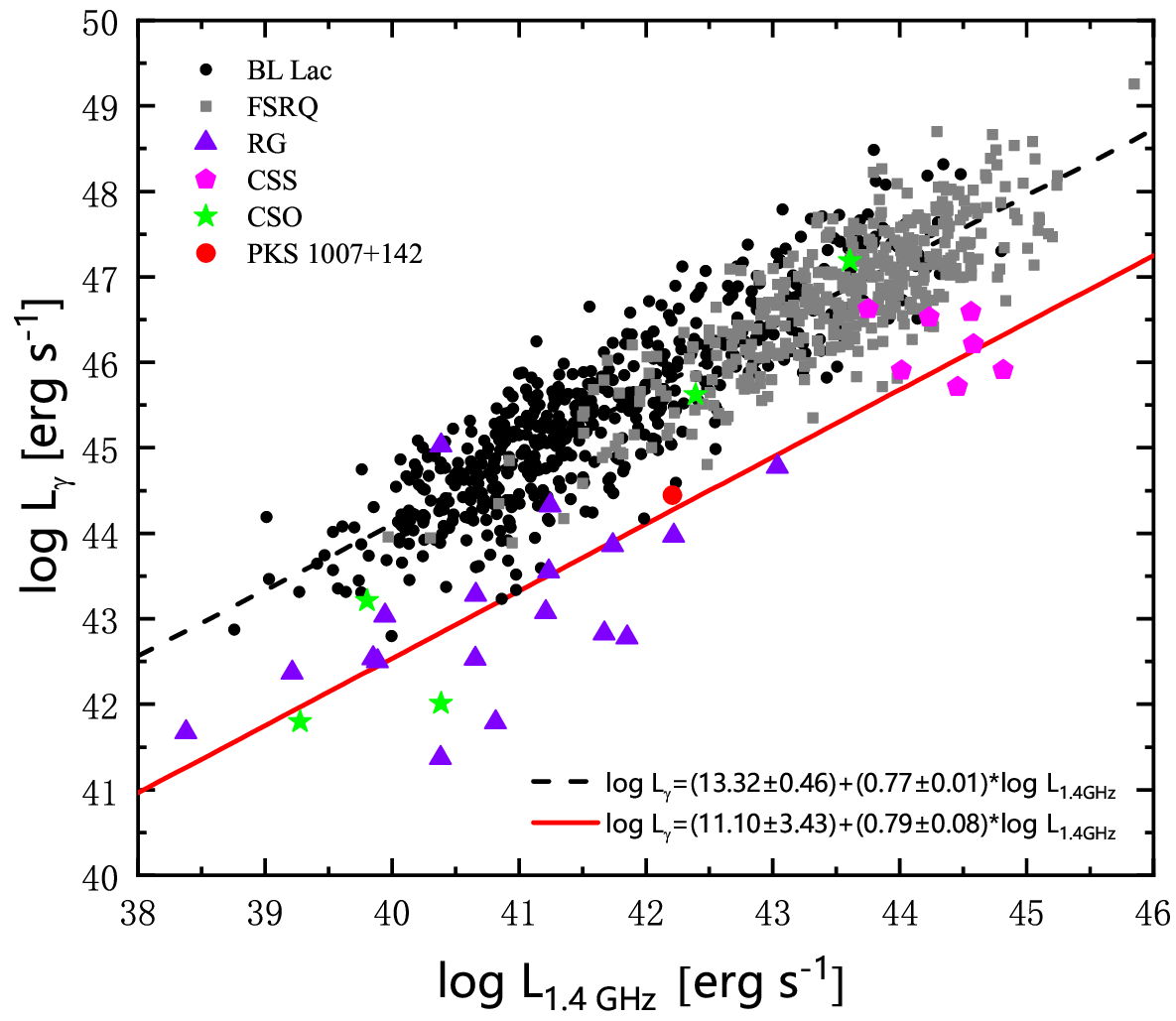}
\caption{\emph{Left panel}: The photon spectral index ($\Gamma_{\gamma}$) vs. the $\gamma$-ray luminosity ($L_{\gamma}$). The data for blazars and radio galaxies (RGs) are obtained from 4FGL-DR3 \citep{2022ApJS..260...53A}, while the data for 7 CSSs are from \cite{2020ApJ...899....2Z} and \cite{2022ApJ...927..221G}, and the data for 5 CSOs are from \citet{2021RAA....21..201G,2022ApJ...939...78G,2024RAA....24b5018G}. \emph{Right panel}: Comparison of PKS 1007+142 with other $\gamma$-ray emitting AGNs in the $L_{\rm 1.4GHz}-L_{\gamma}$ plane. The data at 1.4 GHz of these sources are from the 1.4 GHz NRAO VLA Sky Survey (NVSS, \citealp{1998AJ....115.1693C}), when no NVSS data are available, the data are taken from the NED. The black dashed line represents the linear fitting line for all blazars, while the red solid line is the linear fitting line for RGs, CSSs, and CSOs, excluding PKS 1007+142, CTD 135, and PKS 1413+135.}
\label{Lgamma}
\end{figure}

In order to further reveal the $\gamma$-ray origins of PKS 1007+142, we compare it with other $\gamma$-ray emitting AGNs in the $L_{\gamma}-\Gamma_{\gamma}$ plane (Figure \ref{Lgamma}), where $L_{\gamma}$ is the $\gamma$-ray luminosity. This AGN sample consists of five CSOs (\citealp{migliori2016first,principe2020ngc,principe2021gamma,lister2020txs,2021RAA....21..201G,2022ApJ...939...78G,2024RAA....24b5018G}), seven CSSs \citep{2020ApJ...899....2Z,2022ApJ...927..221G}, together with 18 radio galaxies and numerous blazars \citep[from][]{2022ApJS..260...53A}; blazars includes flat-spectrum radio quasars (FSRQs) and BL Lacertae objects (BL Lacs). According to the criteria of `bona fide' CSOs \citep{2024ApJ...961..240K}, CTD 135 and PKS 1413+135 should be classified as blazar-types \citep[see also][]{2022Symm...14..321F,2022ApJ...927...24P}. In the $L_{\gamma}-\Gamma_{\gamma}$ plane, they are located at the region occupied by blazars, which is distinctly different from the other three `bona fide' CSOs, coinciding with core-jet dominated $\gamma$-rays \citep{2021RAA....21..201G,2022ApJ...939...78G}. In the $L_{\gamma}-\Gamma_{\gamma}$ plane, PKS 1007+142 does not overlap with any other sources. $L_{\gamma}$ of PKS 1007+142 is lower than that of other seven CSSs and FSRQs, higher than three `bona fide' CSOs and most radio galaxies; it is similar to low-luminosity BL Lacs or high-luminosity radio galaxies. These findings align with PKS 1007+142 being classified as a galaxy-type CRS. Notably, PKS 1007+142 almost exhibits the softest spectrum among these $\gamma$-ray emitting AGNs, which may explain its challenging detection by Fermi-LAT. 

In the $L_\gamma-L_{\rm 1.4GHz}$ plane, where $L_{\rm 1.4GHz}$ is the radio luminosity at 1.4 GHz, PKS 1007+142 conforms to the sequence of non-blazars, with a lower ratio of $L_\gamma$ to $L_{\rm 1.4GHz}$ compared to blazars. CTD 135 and PKS 1413+135 still exhibit blazar-like characteristics, whereas the remaining three `bona fide' CSOs are situated within the region of low-luminosity radio galaxies. It is well known that $\gamma$-rays of blazars are predominantly produced by their core jets and are strongly amplified due to the Doppler boosting effect, while their radio radiations originate from more extended regions. These non-blazars have the lower ratios of $L_\gamma$ to $L_{\rm 1.4GHz}$ due to the weakly relativistic effect or the lobe-dominated $\gamma$-ray emission \citep[e.g.,][]{stawarz2008momentum,2010Sci...328..725A,2016ApJ...826....1A,migliori2016first,principe2020ngc,principe2021gamma,2024RAA....24b5018G,2024ApJ...965...163G}. Although it has been confirmed that radio galaxies Cen A, Fornax A, and NGC 6251 are lobe-dominated $\gamma$-ray emitting AGNs \citep{2010Sci...328..725A,2016ApJ...826....1A,2024ApJ...965...163G}, in general, the origin of $\gamma$-rays in radio galaxies is still believed to be from the core-jet \citep{2009ApJ...699...31A,2009ApJ...707...55A,2014A&A...563A..91A,2015ApJ...798...74F,2017RAA....17...90X}. The radio luminosity of PKS 1007+142 surpasses that of most $\gamma$-ray emitting radio galaxies but is lower than that of the seven $\gamma$-ray emitting CSSs. In the $L_\gamma-L_{\rm 1.4GHz}$ plane, PKS 1007+142 exhibits characteristics resembling a high-luminosity radio galaxy in both the radio and $\gamma$-ray bands. 

The recently updated data in the Radio Fundamental Catalog (RFC)\footnote{http://astrogeo.org/rfc/} reveal that PKS 1007+142 exhibits a core-jet structure in its 5 GHz radio image, while the 8 GHz radio image solely displays a core. These observations differ from the 1.6 GHz radio images, which exhibit a `double-lobe' structure without a visible radio core \citep{2010MNRAS.408.2261K}. The luminosity ratio ($R_{\rm CE}$) of the core to extended regions is defined as a core-dominance parameter, $R_{\rm CE}>1$ for core-dominated and $R_{\rm CE}<1$ for lobe-dominated, which is usually taken as an indicator of the beaming effect \citep[e.g.,][]{padovani1992luminosity}. We utilize the 8 GHz radio data from the RFC to estimate the $R_{\rm CE}$ values of these sources. It should be noted that, compared to other $\gamma$-ray emitting radio galaxies, Cen A, Fornax A, and NGC 6251 exhibit relatively low $R_{\rm CE}$ values. For PKS 1007+142, it has $R_{\rm CE}\sim4.2$, similar to CTD 135 and PKS 1413+135 with $R_{\rm CE}\geq1$ at 8 GHz. Conversely, the three `bona fide' CSOs have $R_{\rm CE}<1$. Therefore, we suppose that the $\gamma$-rays of PKS 1007+142 likely originate from its inner jet/core region. However, further observations are required for confirmation or rejection. 

\acknowledgments

The valuable suggestions provided by the anonymous referee are greatly appreciated. We thank the kind permission for the usage of the RFC data from the RFC Collaboration. This research has made use of the NASA/IPAC Extragalactic Database, which is funded by the National Aeronautics and Space Administration and operated by the California Institute of Technology. This work is supported by the National Natural Science Foundation of China (grants 12022305, 12203022, 11973050) and the Natural Science Foundation of Jiangsu Province grant BK20220757.

\clearpage
\bibliography{reference}
\clearpage

\clearpage

\begin{table}[ht!]
\centering
\caption{Location Information of Sources}
\begin{threeparttable} 
\begin{tabular}{lccc}
        \hline\hline
         Name & R.A. & Decl. & References   \\ \hline
        PKS 1007+142 & 152.48\degr & 14.03\degr & \citet{1996AJ....111.1945D} \\ 
        4FGL J1010.0+1416 & $152.52\degr\pm0.04\degr$ & $14.27\degr\pm0.05\degr$ & \citet{ballet2023fermi} \\
        LAT source\tnote{a} & $152.43\degr\pm0.05\degr$ & $14.08\degr\pm0.06\degr$ & \citet{principe2021gamma} \\
        Re-estimated best-fit position & $152.45\degr\pm0.08\degr$ & $14.05\degr\pm0.09\degr$ & This work \\
        \hline\hline
\end{tabular}  
\begin{tablenotes}   
        \footnotesize               
         \item[a] The LAT best-fitting position of PKS1007+142 obtained with 11.3 yr observation data in \citet{principe2021gamma}.
\end{tablenotes}
\end{threeparttable} 
\label{location}
\end{table}

\begin{deluxetable}{cccccc}
\tabletypesize{\scriptsize} \tablecolumns{6} \tablewidth{0pc}
\tablecaption{Information for Each Time Bin in the Light Curve of PKS 1007+142}\tablenum{2}
\label{flux_table}
\tablehead{\colhead{Bin Number} & \colhead{Time\_start} & \colhead{Time\_stop} & \colhead{Flux/Upperlimit} & \colhead{TS} \\ \colhead{} & \colhead{(MJD)} & \colhead{(MJD)} & \colhead{(10$^{-8}$ ph cm$^{-2}$ s$^{-1}$)} & \colhead{}}
\startdata
1 & 54683 & 55048 & 1.39 & 3.9 \\
2 & 55048 & 55413 & 2.01 & 1.9 \\
3 & 55413 & 55778 & 1.93 & 1.2 \\
4 & 55778 & 56143 & 1.74 & 3.8 \\
5 & 56143 & 56508 & 1.29 & 1.2 \\
6 & 56508 & 56873 & 1.22 & 0.1 \\
7 & 56873 & 57238 & 1.79 & 2.1 \\
8 & 57238 & 57603 & 1.12 & 1.2 \\
9 & 57603 & 57968 & 1.94 & 1.3 \\
10 & 57968 & 58333 & 1.54 & 0.3 \\
11 & 58333 & 58698 & 1.72$\pm$0.91 & 15.3 \\
12 & 58698 & 59063 & 2.01$\pm$0.56 & 11.1 \\
13 & 59063 & 59428 & 1.63 & 6.9 \\
14 & 59428 & 59793 & 1.42$\pm$0.59 & 10.2 \\
15 & 59793 & 60158 & 1.12 & 0.5 \\
\enddata
\end{deluxetable}

\clearpage

\end{document}